\documentclass[journal]{IEEEtran}
\ifCLASSINFOpdf

\else

\fi

\usepackage{graphicx,amsmath,amssymb,cite}
\usepackage{subfigure}
\usepackage{color}
\usepackage{amsfonts}
\usepackage{amsmath}
\usepackage{graphicx}
\usepackage{amssymb}
\usepackage{stfloats}
\usepackage{mathrsfs}

\begin{document}
\title{Directional Modulation: A Secure Solution to 5G and Beyond Mobile Networks}
\author{Feng Shu,  Yaolu Qin, Rqing Chen, Ling Xu,  Tong Shen,  Siming Wan\\
Shi Jin, Jiangzhou Wang, and Xiaohu You
\thanks{This work was supported in part by the National Natural Science Foundation of China (Nos. 61771244, 61501238, 61702258, 61472190, and 61271230), in part by the Open Research Fund of National Key Laboratory of Electromagnetic Environment, China Research Institute of Radiowave Propagation (No. 201500013), in part by the Jiangsu Provincial Science Foundation under Project BK20150786, in part by the Specially Appointed Professor Program in Jiangsu Province, 2015, in part by the Fundamental Research Funds for the Central Universities under Grant 30916011205, and in part by the open research fund of National Mobile Communications Research Laboratory, Southeast University, China (Nos. 2017D04 and 2013D02).}
}

\maketitle

\begin{abstract}
Directional modulation (DM), as an efficient secure transmission way, offers security through its directive property and is suitable for line-of-propagation (LoP) channels such as millimeter wave (mmWave) massive multiple-input multiple-output (MIMO), satellite communication, unmanned aerial vehicle (UAV), and smart transportation. If the direction angle of the desired received is known, the desired channel gain vector is obtainable. Thus, in advance, the DM transmitter knows the values of directional angles of desired user and eavesdropper, or their direction of arrival (DOAs) because the beamforming vector of confidential messages and artificial noise (AN) projection matrix is mainly determined by directional angles of desired user and eavesdropper. For a DM transceiver, working as a receiver, the first step is to measure the DOAs of desired user and eavesdropper. Then, in the second step, using the measured DOAs, the beamforming vector of confidential messages and AN projection matrix is designed. In this paper, we describe the DOA measurement methods, power allocation, and beamforming in DM networks. A machine learning-based DOA measurement method is proposed to make a substantial SR performance gain compared to single-snapshot measurement without machine learning for a given null-space projection beamforming scheme. However, for a conventional DM network, there still exists a serious secure issue: the eavesdropper moves inside the main beam of the desired user and may intercept the confidential messages intended to the desired users because the beamforming vector of confidential messages and AN projection matrix are only angle-dependence. To address this problem, we present a new concept of secure and precise transmission, where the transmit waveform has two-dimensional even three-dimensional dependence by using DM, random frequency selection, and phase alignment at DM transmitter.
\end{abstract}

\section{Directional modulation: concept and system}
Due to the broadcast nature of wireless transmission, confidential information is easily intercepted by malicious users, which will expose wireless networks to serious secure risk. Therefore, secure transmission, storage, processing, and protection of confidential information become extremely important research topics in wireless networks in recent years. Additionally, there is an increasing demand to design secure transmission methods to protect the legitimate users from being overheard. Recently, as millimeter wave (mmWave) massive multiple-input multiple-output (MIMO), unmanned aerial vehicle (UAV), and internet of things (IoT)  emerge, how to securely transmit confidential messages in line-of-propagation (LoP) channel is becoming a hot research area. Directional modulation (DM),  as an inherent secure method in LoP channel, attract more and more research activities from both academia and industry communities.

\begin{figure}[h]
\centering
\includegraphics[width=8cm]{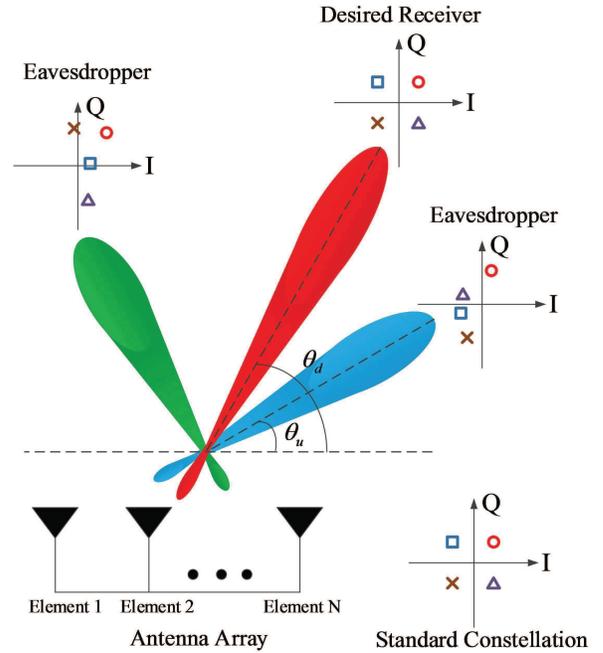}\\
\caption{Schematic diagram of directional  modulation network.}\label{DM_Network}
\end{figure}

\begin{figure*}[htbp]
\centering
\includegraphics[width=16cm]{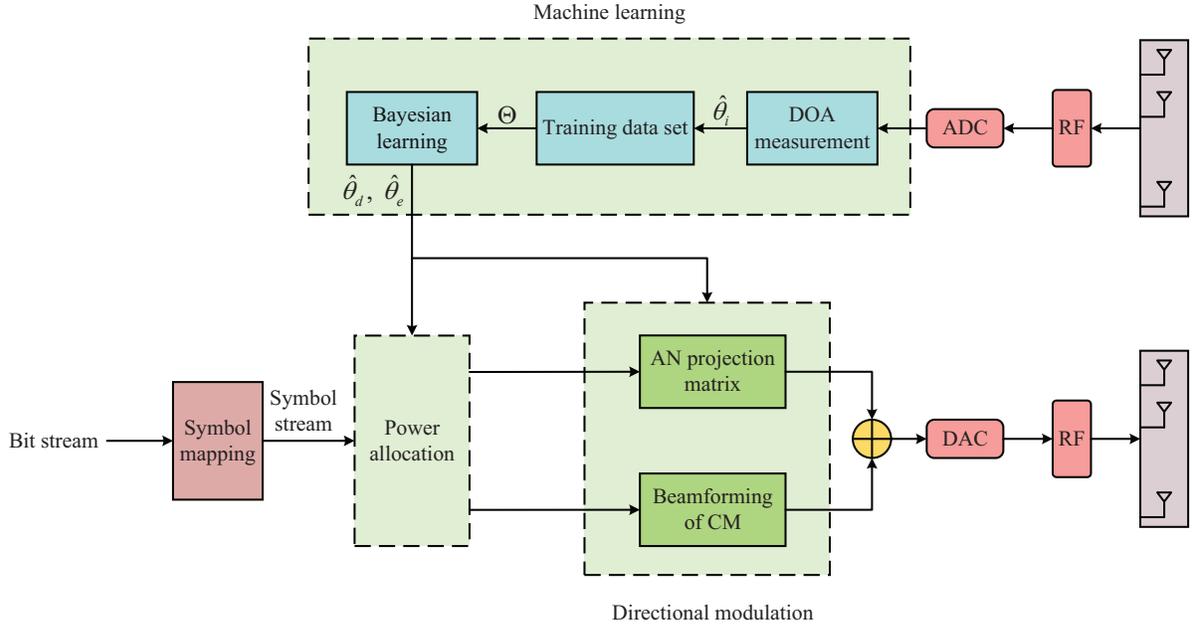}
\caption{Schematic diagram of smart directional modulation transceiver, where CM stands for confidential messages.}\label{Smart_DM_TxRx}
\end{figure*}

In Fig.~\ref{DM_Network}, a concept diagram is presented to show the basic idea of DM. In this figure, there are one DM transmitter with multiple transmit antennas, two eavesdroppers, and one desired user. The DM transmitter sends its confidential messages to the desired user such that the two eavesdroppers cannot intercept the confidential messages. By exploiting the directive property generated by antenna-array-based beamforming with the help of artificial noise (AN) projection, the DM transmitter transmits the confidential message to the desired direction and interferes with the eavesdroppers by AN projection. Due to the joint operation of AN projection and beamforming, the confidential messages can be securely and successfully transmitted to the desired user, and AN severely corrupts the eavesdroppers. We propose a smart DM transceiver shown in Fig.~\ref{Smart_DM_TxRx}. Here, one of machine learning methods, Bayesian learning in \cite{Bishop} is adopted to improve the measurement precision of DOA angles, which subsequently serves for designing the precoding vector and AN projection matrix at DM. Due to channel noise and interference in channel, there exists the DOA measurement errors. As shown in \cite{Hu2016}, a high-precision DOA estimation means a large secrecy rate provided that a fixed beamforming scheme is given. In what follows, the number of antennas at DM transmitter is assumed to be $N$.

\section{DOA measurement,  and Machine learning}

For secure DM,   high-resolution DOA measurement and high-precise  DOA error density estimation  are significantly important for DM transmitter.  As stated in \cite{Hu2016}, \cite{Xu1},
perfect DOA measurement results in performance improvement since expected information and AN can be accurately transmitted to the desired direction and eavesdropping direction, respectively. Due to ultra-high-resolution of spatial direction, and super-high-spectral efficiency, massive MIMO has drawn tremendous research attention from academia and industry communities \cite{Larsson}. If massive MIMO behaves as a receive array, DOA estimation precision will be dramatically improved due to its ultra-high-resolution of spatial direction.

The estimated signals can be classified into two categories: narrowband signals and wideband signals \cite{Tuncer}. The narrowband signals are transmitted with the same frequency. However, in wideband signal DOA estimation problems, several frequency bins carry different information corresponding to source DOA angles. This generates diversity and processing gain for DOA estimation. The main problem in wideband DOA estimation is how the information at each frequency can be combined to obtain the most accurate DOA information. Two main types of wideband processing are coherent and noncoherent. In coherent wideband processing, the covariance matrix at each frequency is combined coherently using a transformation and the final covariance matrix is used for DOA estimation. In noncoherent wideband processing, the DOA at each frequency is estimated separately and the results are combined noncoherently.

Many typical DOA estimation algorithms have been proposed and analyzed. Capon algorithm is maximum likelihood estimation of power which aims at maximizing the signal-to-interference ratio (SINR). Schmidt developed a more popular method, i.e., the multiple signal classification (MUSIC) \cite{Schmidt} algorithm, which is a high-resolution eigen-structure-based DOA-finding method.


Considering the need of high-resolution DOA measurement in physical layer secure communication and the advantages of massive MIMO systems \cite{Larsson}, adopting massive MIMO in DOA measurement is reasonable. However, as the number of antennas tends to be large-scale, the computational amount, circuit complexity and cost of  digital implementation become too high for commercial applications. Therefore, a hybrid analog and digital (HAD) structure is a natural choice, which will strike a good balance among computational amount, circuit cost, and circuit implementation complexity. In \cite{Qin}, the authors presented the low-complexity DOA estimation algorithm based on Root-MUSIC method and hybrid structure which can reduce both the computational complexity and hardware cost.

%

\begin{figure}[h]
\centering
\includegraphics[width=0.5\textwidth]{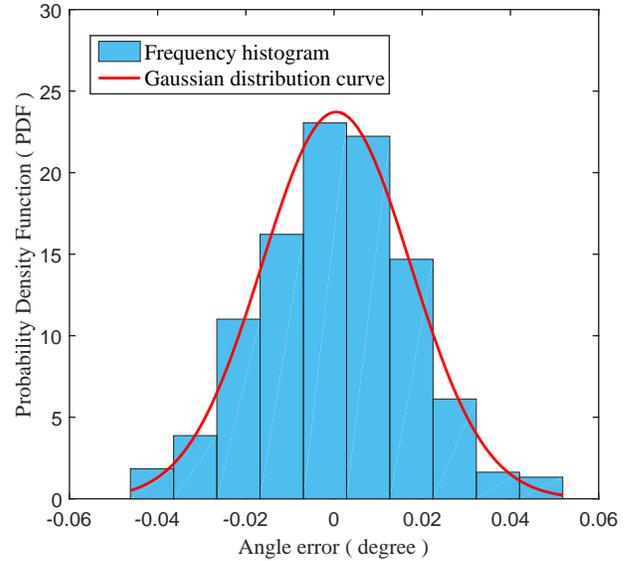}\\
\caption{Histogram of DOA measurement errors.}\label{fig3_DOA_Hisgram}
\end{figure}

In a DM network, the beamforming vector and AN projection matrix depend heavily on the precision of DOA measurement. How to achieve a high-precision DOA measurement, machine learning is a new and advanced method to address such a problem. In this article, we present a Bayesian learning (BL) based method to improve the DOA measurement precision at DM transmitter  with fully-digital (FD) structure. It is particularly noted that the idea can be readily extended to the HAD structure. Firstly, the DM transmitter  with FD structure  acts as  a receiver, and takes a number of snapshots.  By using the proposed method Root-MUSIC, we can obtain a training data set (TDS) of measured DOA values by taking multiple snapshots and its size is denoted as $K$, where the number of sampling points is $M$. As shown in Fig.~\ref{fig3_DOA_Hisgram},  the histogram method in machine learning is adopted to do density estimation of DOA measurement errors. In accordance with this figure, it is clear that the DOA measurement errors approximately obeys the Gaussian distribution. Subsequently, the DOA measurement error is viewed as a Gaussian distribution in the following. The BL is introduced to estimate the mean and variance of the desired DOA measurement error, where the prior distributions of angle mean and variance are assumed to be uniformly distributed.

\begin{figure}[h]
\centering
\includegraphics[width=0.5\textwidth]{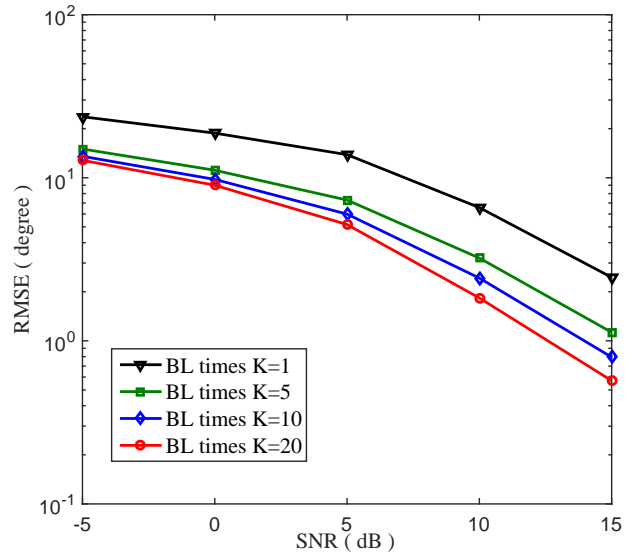}\\
\caption{Root mean squared error of BL-based DOA measurement versus SNR($M=4$).}\label{fig4_DOA_RMSE}
\end{figure}

Fig.~\ref{fig4_DOA_RMSE}  illustrates the curves of root mean squared error (RMSE) of DOA measurement versus  signal-to-noise ratio (SNR) for BL with different sizes of training data set, where the size of TDS is $K$. It is very clear that as the size $K$ of TDS increases from 1 to 20, the DOA measurement precision is improved gradually and obviously.

\begin{figure}[h]
\centering
\includegraphics[width=0.5\textwidth]{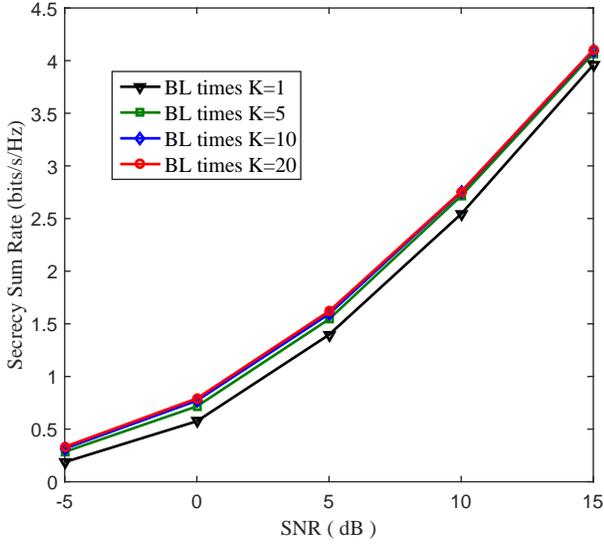}\\
\caption{Secrecy rate versus SNR with different sizes of training data set($M=4$).}\label{fig5_SR}
\end{figure}

Fig.~\ref{fig5_SR} demonstrates the curves of  secrecy rate (SR) versus  signal-to-noise ratio (SNR) for in a DM network as shown in Fig.~\ref{DM_Network} , where null-space projection (NSP) method is adopted to design AN projection matrix and matched filter (MF) is for the precoding vector of confidential message, the DOA values  measured  by BL is directly substituted into the expressions of NSP injection and MF beamforming. As shown in Fig.~\ref{fig5_SR}, it is seen that the achievable SR performance gain  achieved by increasing the size $K$ of TDS  is appreciated, especially in the low SNR region.  However, as SNR increases, the gain will decrease slowly and gradually. It is particularly noted that the DOA measurement for desired user and eavesdropper is done by receiving the transmit signal from desired user and eavesdropper in the first time slot once the DM transceiver behaves as a receiver. In the second time slot, via the beamforming scheme, the DM transceiver operates in transmit model,  sends the confidential messages to the desired user, and projects AN towards the eavesdropper.

\section{Power allocation, robust beamforming, and AN projection for DM}

As an extremely important secure physical layer wireless transmission way, DM can enable confidential messages transmit along  a specified direction. In particular, DM is a transmitter side technology that projects digitally modulated signals into a pre-specified safe direction  and simultaneously distorts the constellation formats of the signals in other directions by AN projection. The traditional DM technology is mainly investigated in the radio frequency front end. One kind of DM transmitters that rely on near-field coupling effects, and this kind of DM transmitters can be regarded as passive DM transmitters because all parasitic elements are passively excited. In this case, the combinations of each switch state on near-field passive reflectors can form a unique far-field radiation pattern that can be converted into detected constellation points in IQ space along each spatial direction in free space. After the tremendous measurements of multiple modes for a large number of possible switch states combinations, the available constellation patterns in the desired direction can be selected and the corresponding switch settings for secure transmission of each symbol can be obtained. However, constellations along other unselected directions are distorted in a complex and almost unpredictable way due to the complicated  passive near-field diffraction effect along spatial direction. Besides, the complex interaction in the near field and their spatially correlated transformation with the far field make the synthesis of passive DM transmitters immensely difficult.

Daly first proposed a method to simplify DM synthesis in \cite{MPDaly1}, in which the structures of the DM transmitters consist of actively driven antenna arrays and is equipped with a reconfigurable phase shifter or radiator. Since each array element is actively excited, this kind of DM transmitters is called active DM array. In active DM arrays, the carrier signal needs to be modulated by the baseband information data which controls the attenuator and the phase shifter before being transmitted through the antenna array. In general, this type of DM transmitter is synthesized by minimizing the value of an appropriate cost function that sets the architecture parameters and predicts system performance through iterative optimization. Based on the same physical structure, some parameters such as quantization phase shifter, array element spacing, and active element mode are studied exhaustively, respectively.

Another way to implement DM is performed via baseband signals. It is notable that the author in \cite{YDing} proposed a new theory that describes direction modulation system with the expression of vector, and with the help of this theory, the transmission characteristics of DM was achieved. This synthesis method is also known as the orthogonal vector method. Compared with the traditional synthesis method that implemented DM at RF front-end, this synthesis method can be easily implemented in baseband signal by utilizing the beamforming scheme and added AN, and in this case, it can further ensure that different constellation points are transmitted in different time slots, in other words, the dynamic DM. However, all these works are based on the perfect situation of the desired direction angle, and they do not take the estimation error of desired direction angle in actual scenarios into consideration.
\begin{figure}[htp]
\centering
\subfigure[$\beta=0.2$]{
\includegraphics[width=0.48\textwidth]{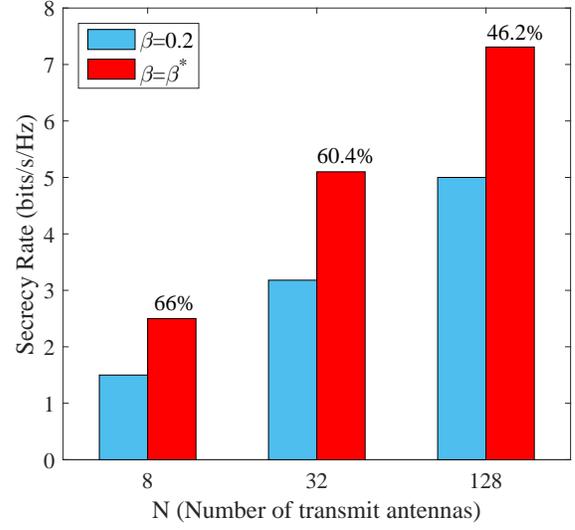}}
\hspace{1in}
\subfigure[$\beta=0.5$]{
\includegraphics[width=0.48\textwidth]{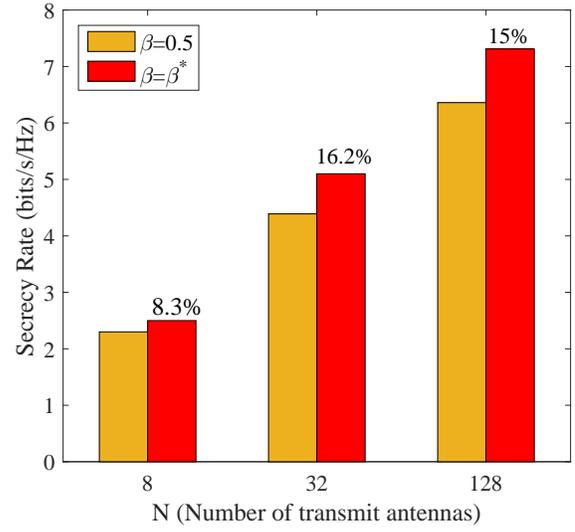}}
\hspace{1in}
\subfigure[$\beta=0.8$]{
\includegraphics[width=0.48\textwidth]{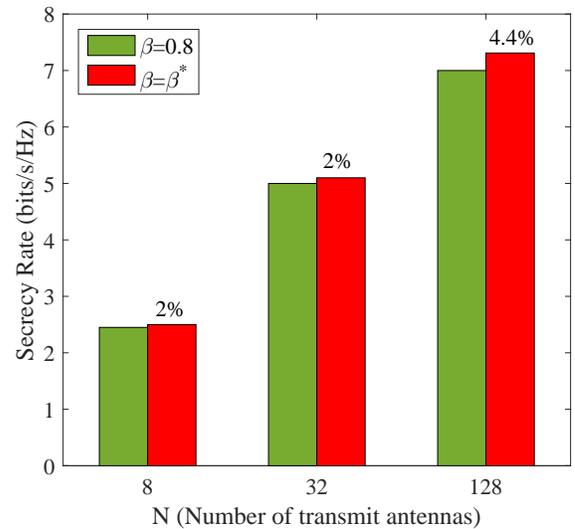}}
\caption{Histogram of Secrecy rate versus number of antennas at DM transmitter.}\label{Fig6_Perf}
\end{figure}

%
%
%
%

In the recent several years, the DOA measurement error is exploited and some robust beamforming methods of designing the beamforming vector of confidential messages and projection matrix of AN were proposed \cite{Hu2016,Shu2016,Zhu}. The authors of \cite{Hu2016} derived a closed-form expression for the null space of the steering vector along the desired direction, and  a robust DM synthesis method based on conditional minimum mean square error (MMSE) criterion was proposed to minimize the distortion of constellation points along the legitimated users. Subsequently, in \cite{Shu2016}, the authors extended the idea of \cite{Hu2016} to multi-beam DM scenarios in broadcasting systems, where a conditional maximizing signal-to-leakage noise ratio (Max-SLNR) was presented to design the beamforming vector of confidential messages and maximizing the signal-to-artificial-noise ratio (Max-SANR) at the desired receivers was employed to design projection matrix of AN. However, the two robust schemes\cite{Hu2016,Shu2016} need to know the PDF of DOA measurement error in advance. This imposes a great burden on DOA measurement operation. To address this problem , without the knowledge of distribution of direction measurement error, a blind robust scheme of combining main-lobe-integration and leakage were formed to design the beamforming vector of confidential messages and AN projection matrix in multi-user multiple input multiple output (MIMO) scenario. Additionally, the authors extend the application of DM to a multi-cast scenario in \cite{Xu1}. In the extended work, two novel schemes,  maximum group receive power plus null-space projection scheme and Max-SLNR plus maximum-AN-leakage-and-noise ratio scheme, were proposed  to improve the secrecy sum-rate.

How does the power allocation strategy between AN and confidential messages affect?  In \cite{WSM2018}, given a fixed beamforming scheme,  an optimal power allocation (OPA) strategy of maximizing secrecy rate  for secure directional modulation is presented and its closed-form formula is also derived. In Fig.~\ref{Fig6_Perf}, the SR performance gain percentage achieved by OPA over three fixed typical power allocation methods are demonstrated, respectively. Here, $\beta$ stands for the PA factor for confidential messages. Observing the three figures, it is very obvious that a small value $\beta$ means a small performance improvement. In small-scale antenna array and low SNR region, the achievable SR performance attained by OPA is relatively attractive. For example, at $N=8$, $\beta=0.2$, and SNR=5dB, the SR performance gain percent is about $66\%$. This improvement is very impressive.

\section{Secure Precise Wireless Transmission}
  For a conventional DM network, if an eavesdropper moves inside the mainbeam of the desired user and even has different distance to the DM transmitter compared to the desired user, then it is still intercept the confidential messages for the desired user. This leads to a serious secure issue.  To address the serious security problem faced by traditional DM,   a new concept, called secure and precise wireless transmission, was first proposed by the authors in \cite{Hu2016Artificial} and \cite{Shu2017Secure}, where three key techniques, DM, random frequency diversity, and phase alignment,  are utilized to reach the secure and precise transmission,  which generates two-dimensional angle and distance dependence.  Now, there is another way to achieve secure and precise wireless transmission: multi-relay cooperation with the aid of DM\cite{Zhu2017}. In other words, given angle and range,  the confidential message power can be transmitted  and collected inside  a small neighborhood around the desired position, outside which there exists very weak receive power seriously corrupted by AN and the confident messages can not be detected successfully.

In \cite{Hu2016Artificial}, a random frequency diversity array was proposed which randomly shifts the carrier frequencies in each different transmit antennas, and the frequency shifting on each element has an independent and identically distributed statistical property, decoupling the correlation between the distance and the direction angle while guaranteeing the array's dependence to the distance and direction angle. In addition, the transmitter added AN to interfere with other location of the eavesdropper, so that this scheme can significantly improve the ergodic safety capacity of information transmission, and improve the safety transmission performance.

 Another secure precise transmission scheme with multi-relay-aided directional modulation was also proposed in \cite{Zhu2017}. In this scheme, confidential messages are transmitted by multiple relays, and utilize direction modulation on every relay. All relays adjusted their directive main beams to the desired position such that the power peak is formed by coherent superposition. Deviation from this position, the receive confidential power add together not constructively perhaps destructively. Furthermore, the combined signal is also seriously corrupted by AN.

Due to the use of MIMO array and random frequency selection at transmitter,  where $N$ denotes the number of transmit antennas at DM transceiver, the desired receiver requires $N$ RF chains to coherently combine all signals from $N$ frequency bands. As $N$ tends to large scale,  the circuit cost and computational complexity at desired receiver will increase linearly and dramatically. To reduce this cost, in \cite{Shu2017Secure},  a secure and precise wireless transmission with random subcarrier selection (RSS) based on OFDM and direction modulation  was proposed. This scheme  reduces  $N$ RF chains to only one. At receiver, via the FFT/IFFT operation at basedband,   in the frequency domain the phase alignment is realized readily.  Although the OFDM-based RSS may dramatically reduce the circuit cost and computational complexity of the receiver, the DM transmitter still keeps a high circuit cost. A HAD structure at DM transmitter is a natural alternative.

%
%

\begin{figure}[htp]
\centering
\subfigure[SINR versus  elevation angle and distance]{
\includegraphics[width=0.5\textwidth]{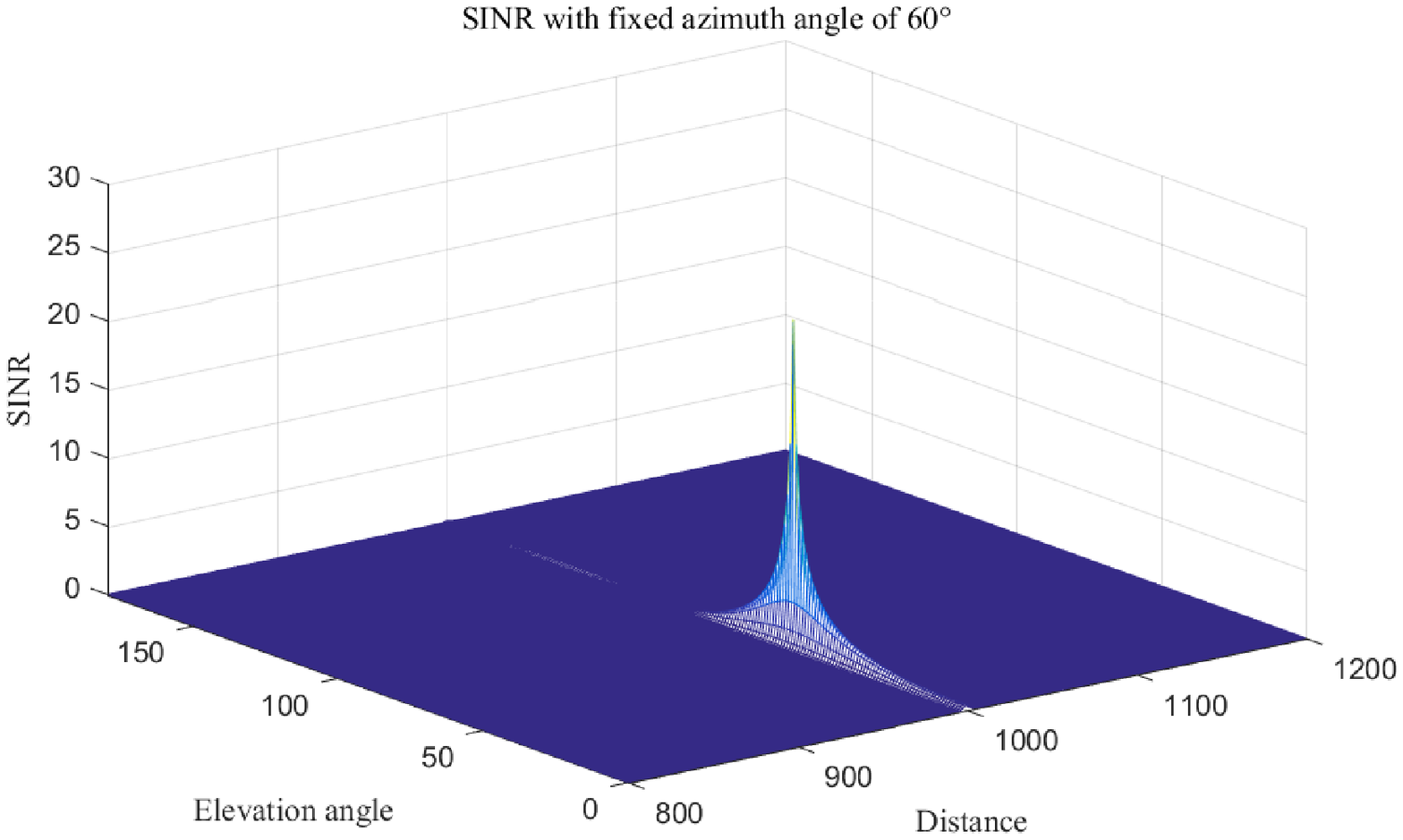}}
\hspace{1in}
\subfigure[SINR versus  azimuth angle and distance]{
\includegraphics[width=0.5\textwidth]{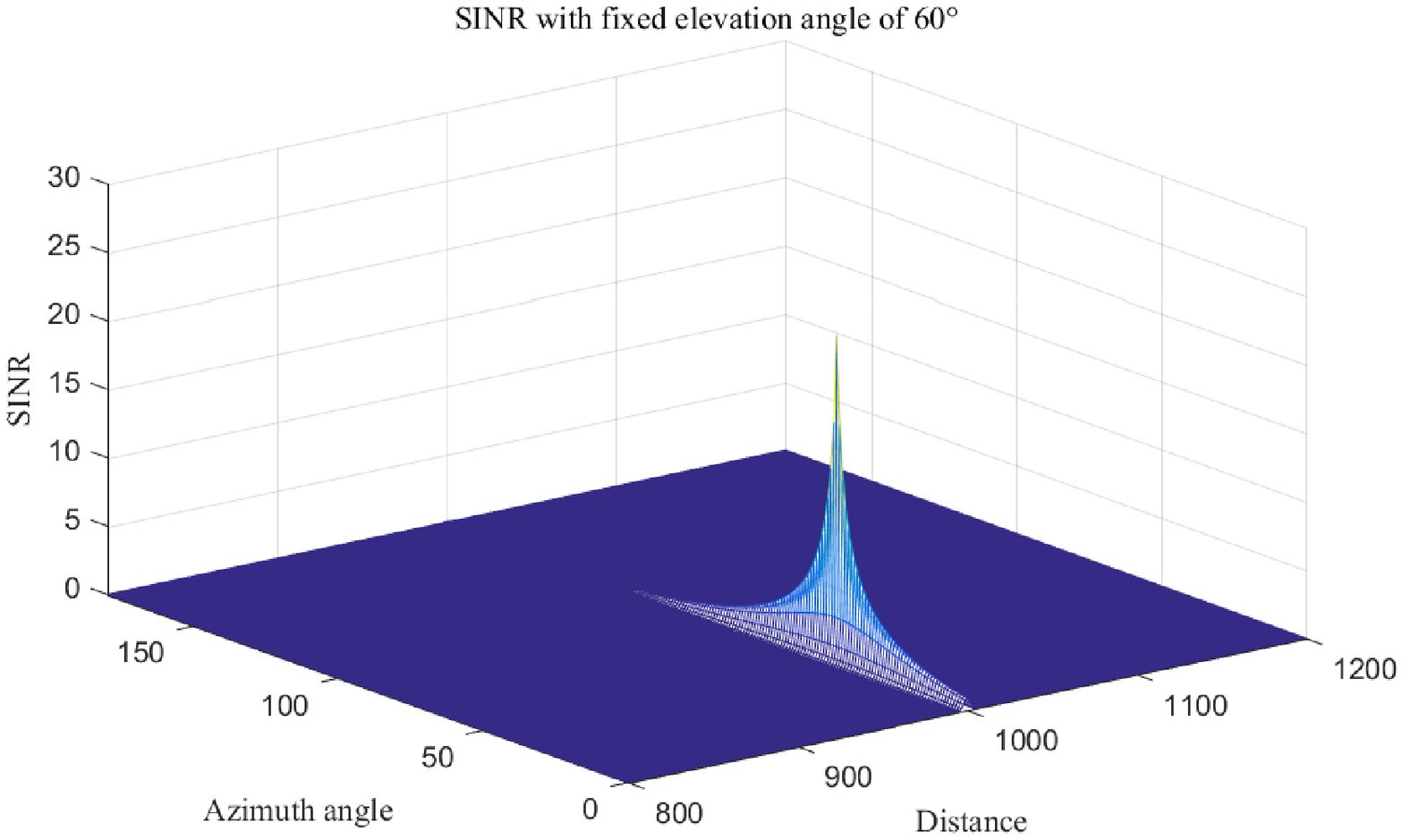}}
\hspace{1in}
\subfigure[SINR versus azimuth  and elevation angles]{
\includegraphics[width=0.5\textwidth]{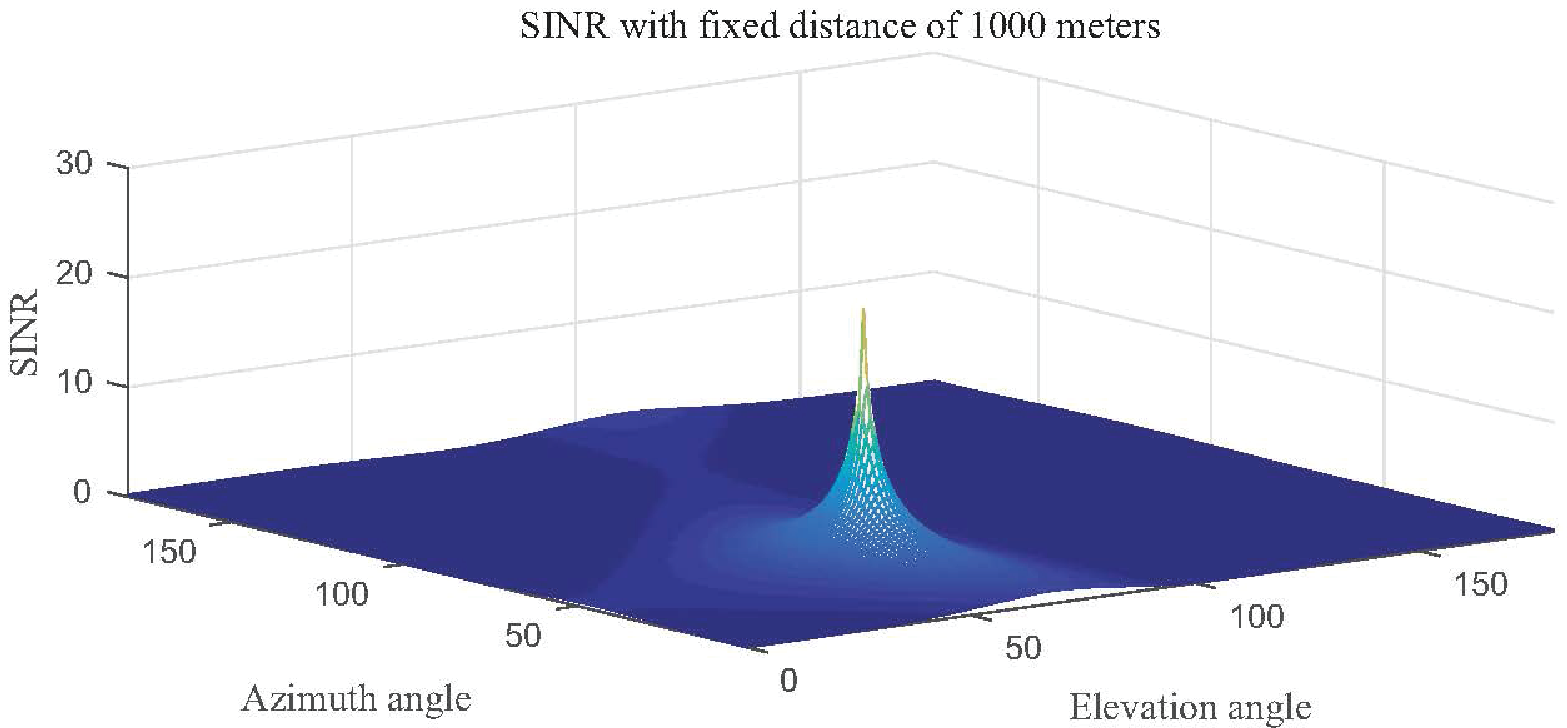}}
\caption{Average SINR  over: (a) elevation angle and distance, (b) azimuth angle and distance, and (c) azimuth and elevation angles,   where $B=5$MHz.}\label{fig7}
\end{figure}
In the above secure and precise wireless system \cite{Hu2016Artificial,Shu2017Secure,Zhu2017}, only two-dimensional (2D) azimuth angle and distance were exploited.  In reality, transmitter and receiver can not be guaranteed to be in the same plane.  How to achieve precise wireless transmission in three-dimensional (3D) scene?  However, a practical environment is 3D: elevation angle, azimuth angle and distance. It is also an interesting work to extend traditional 2D DM to 3D DM. For a 3D DM, the beamforming scheme, including AN projection matrix and confidential beamforming vectors in \cite{Shu2017Secure}, is  extended to the 3D polar coordinate system. Fig.~\ref{fig7} shows the average SINR over any two coordinates with the third coordinate being fixed, where the desired and eavesdropper  directions are 45$^{\circ}$, and 120$^{\circ}$. It is seen that only single SINR peak appears  in all three subfigures (a), (b), and (c). Outside the single peak, there are only weak confidential message power seriously corrupted by AN.  This confirms the fact that the secure and precise transmission actually addresses the secure problem of DM, where the traditional DM as shown in \cite{Hu2016Artificial} has the confidential message power mountain stretching from the DM transmitter to infinity easily intercepted by those eavesdroppers who move inside the main beam of the desired user.


%

Due to the above advantages of DM and its related improved version such as secure and precise transmission, DM may extend its application fields to more aspects.  However, there still exist so many open problems to address in secure and precise transmission, and DM fields. Here, we list multiple important ones of them as follows:

\begin{enumerate}
  \item All above schemes only consider single desired user, How to extend the above secure and precise wireless transmission  to multi-user, broadcasting, and multi-cast scenarios. In such a situation, the DM
  transmitter need to generate multiple peaks of confidential message. This will place a great challenge on how to design a set of random subcarriers, who forms multiple power peaks corresponding to multiple desired receivers
   \item As the number of transmit antennas at DM transmitter tends to large-scale, the circuit cost and complexity becomes a significant obstacle for practical applications of DM. A FD structure is clearly infeasible. A HAD beamforming structure is a natural choice. Sub-array structure in HAD can result in a dramatic reduction in circuit cost and complexity. However, in such a structure, there are open problems of how to offer high-precision DOA measurement and robust beamforming scheme for DM.

    \item In wireless mobile communication, the relative motion  among Alice, Bob,  and Eve will produce the Doppler shift. How to measure the DOA  under Doppler shift is a  challenging task.  Even how to jointly estimate Doppler shift and DOA is also a new problem.

    \item DM is originally suitable for LoP channels. In multipath channel, multipath effect will gather AN towards the desired receiver if the blocking object locates at the eavesdropper direction. This effect is called the effect of gathering AN due to multipath transmission. It will severely degrade the performance of the desired receiver. How to combat this effect is also a major challenge for DM. If this problem is solved, DM can extend its applications to multipath channel, and even significantly enlarge its application fields.
\end{enumerate}

\section{Conclusion}

In this article,  the great potential of directional modulation has been highlighted as a key enabling secure technology for future fifth generation (5G) cellular systems, IoT, UAV, smart transportation, and satellite communications. We review its beamforming schemes, DOA estimation, power allocation, and its improved version secure and precise transmission. The technology achieves  excellent security  in LoP channels. There are still several challenging problems ahead to exploit the full potential of the technology. Also, we have raised several new open important research problems. Finally, in our view, DM will achieve  wide diverse promising applications in the near future.
\ifCLASSOPTIONcaptionsoff
  \newpage
\fi

\bibliographystyle{IEEEtran}
\bibliography{IEEEfull,cite}

\begin{IEEEbiographynophoto}\\
FENG SHU is a professor with the School of Electronic and Optical Engineering at Nanjing University of Science and Technology, Nanjing, China, also with School of Computer and  information at Fujian Agriculture and Forestry University and awarded with Mingjian Scholar Chair Professor in Fujian Province, China. He has published more 200 journal and conference papers on signal processing and
communications. His research interests include statistical signal processing and low-complexity algorithms with applications in wireless communications.
\end{IEEEbiographynophoto}
\begin{IEEEbiographynophoto}\\
YAOLU QIN is a postgraduate student in the School of Electronic and Optical Engineering at Nanjing University of Science and Technology, Nanjing, China. Her research interests include massive MIMO, wireless localization, and DOA measurement in wireless communications.
\end{IEEEbiographynophoto}
\begin{IEEEbiographynophoto}\\
RIQING CHEN is a professor with School of Computer and  information at Fujian Agriculture and Forestry University, Fujian Province, China.  He has published more 50 journal and conference papers on signal processing and communications. His research interests include statistical signal processing and low-complexity algorithms with applications in wireless communications.
\end{IEEEbiographynophoto}
\begin{IEEEbiographynophoto}\\
TONG SHEN is a PhD student in the School of Electronic and Optical Engineering at Nanjing University of Science and Technology, Nanjing, China. His research interests include massive MIMO, directional modulation, and DOA measurement in wireless communications.
\end{IEEEbiographynophoto}
\begin{IEEEbiographynophoto}\\
LING XU is a postgraduate student in the School of Electronic and Optical Engineering at Nanjing University of Science and Technology, Nanjing, China. Her research interests include massive MIMO, wireless localization, and DOA measurement in wireless communications.
\end{IEEEbiographynophoto}
\begin{IEEEbiographynophoto}\\
SIMING WAN is a postgraduate student in the School of Electronic and Optical Engineering at Nanjing University of Science and Technology, Nanjing, China. Her research interests include massive MIMO, wireless localization, and DOA measurement in wireless communications.
\end{IEEEbiographynophoto}
\begin{IEEEbiographynophoto}\\
SHI JIN is a professor School of Information Science and Enginering  at Southeast University, Nanjing, China.  He has published more 100 journal and conference papers on signal processing and communications. His research interests include statistical signal processing and low-complexity algorithms with
applications in wireless communications.
\end{IEEEbiographynophoto}
\begin{IEEEbiographynophoto}\\
JIANGZHOU WANG is currently the Head of the School of Engineering and Digital Arts and a Professor of Telecommunications with the University of Kent, Canterbury, U.K. He has authored over 200 papers in international journals and conferences in the areas of wireless mobile communications and three books. His research interests include statistical signal processing and low-complexity algorithms with applications in wireless communications.
\end{IEEEbiographynophoto}
\begin{IEEEbiographynophoto}\\
XIAOHU YOU is with Southeast University, first as an Associate Professor and later as a Professor. He is also the Chief of the Technical Group of China 3G/B3G Mobile Communication R\&D Project. His research interests include mobile communications, adaptive signal processing, and artificial neural networks with applications to communications and biomedical engineering.
\end{IEEEbiographynophoto}
\end{document}